\documentclass[11pt]{article}
\usepackage{graphicx}
\usepackage{amssymb}
\usepackage{amsmath}
\usepackage{graphicx}
\usepackage{amstext}
\usepackage{esint}
\usepackage[utf8]{inputenc}
\usepackage{color}
\usepackage{float}
\usepackage{cite}

 \numberwithin{equation}{section}

\def\beq{\begin{eqnarray}}    
\def\eeq{\end{eqnarray}}      

\newcommand{\wBD}{\omega_{\rm BD}}
\newcommand{\eBD}{\epsilon_{\rm BD}}

\begin{document}

 \hyphenation{nu-cleo-syn-the-sis u-sing si-mu-la-te ma-king
cos-mo-lo-gy know-led-ge e-vi-den-ce stu-dies be-ha-vi-or
res-pec-ti-ve-ly appro-xi-ma-te-ly gra-vi-ty sca-ling
ge-ne-ra-li-zed re-mai-ning va-cu-um}

\begin{center}
{\bf \Large Difficulties in reconciling non-negligible differences between the local and cosmological values of the gravitational coupling in extended Brans-Dicke theories}\vskip 2mm

 \vskip 8mm

\textbf{\large Adri\`a G\'omez-Valent and Prajwal Hassan Puttasiddappa}

\vskip 0.5cm
Institut f\"{u}r Theoretische Physik, Ruprecht-Karls-Universit\"{a}t Heidelberg,
Philosophenweg 16, D-69120 Heidelberg, Germany

\vskip0.3cm

{E-mails: gomez-valent@thphys.uni-heidelberg.de, puttasiddappa@thphys.uni-heidelberg.de}

 \vskip2mm

\end{center}
\vskip 7mm

\begin{quotation}
\noindent {\large\it \underline{Abstract}}.
Recent studies by Sol\`a Peracaula, G\'omez-Valent, de Cruz P\'erez and Moreno-Pulido (2019,2020) have pointed out the intriguing possibility that Brans-Dicke cosmology with constant vacuum energy density (BD-$\Lambda$CDM) may be able to alleviate the $H_0$ and $\sigma_8$ tensions that are found in the framework of the concordance cosmological model (GR-$\Lambda$CDM). The fitting analyses presented in these works indicate a preference for values of the effective gravitational coupling appearing in the Friedmann equation, $G$, about $4-9\%$ larger than Newton's constant (as measured on Earth), and mildly evolving with the expansion of the universe. The signal reaches the $\sim 3.5\sigma$ c.l. when the prior on $H_0$ from SH0ES and the angular diameter distances to strong gravitationally lensed quasars measured by H0LICOW are considered, and the $\sim 3\sigma$ c.l. when only the former is included. Thus, the improvement in the description of the cosmological datasets relies on the existence of a mechanism capable of screening the modified gravity effects at those scales where deviations from standard General Relativity (GR) are highly constrained, as in the Solar System. In this paper we explore several extensions of BD-$\Lambda$CDM that can leave the cosmological evolution basically unaltered at the background and linear perturbations level, while being able to screen the Brans-Dicke effects inside the regions of interest, leading to standard GR. We search for weak-field solutions around spherical static massive objects with no internal pressure and show that, unfortunately, these mechanisms can only explain very tiny departures of the effective cosmological gravitational coupling from the one measured locally. This might hinder the ability of BD-$\Lambda$CDM to alleviate the cosmological tensions.   

\end{quotation}
\vskip 3mm

\begin{quotation}
\end{quotation}

\newpage


\newpage



\section{Introduction}\label{intro}

The standard model of Cosmology, also known as concordance or $\Lambda$CDM model, is constructed on the basis of General Relativity and the presence of a positive cosmological constant in the action, $\Lambda$, that accompanies the usual Einstein-Hilbert term \cite{Einstein1917}. This constant has an associated energy density $\rho_\Lambda=\Lambda/8\pi G_N$ which is of course also constant in time and homogeneous. The latter behaves effectively as a vacuum energy density with gravitational repulsive effects that pervades the universe and is in charge of its current positive acceleration. The concordance model also relies (among other things) on: (i) the symmetries of the Cosmological Principle; (ii) the existence of an inflationary phase of exponentially fast expansion previous to the radiation-dominated era that is needed to solve the horizon, flatness and magnetic monopole problems, and also to set the seeds of the structures that we observe nowadays in the universe; and (iii) the presence of cold dark matter (CDM) to allow for the efficient formation of the aforesaid structures. Although there exist important theoretical conundrums related with some of the fundamental blocks of the model, as e.g. the famous old cosmological constant problem \cite{Weinberg1989,JeromeMartin2012,Sola2013}, the concordance model has been able to explain a large variety of cosmological observations, namely the abundances of light elements \cite{Steigman2007}, the cosmic microwave background (CMB) anisotropies \cite{Planck2018}, the baryon acoustic oscillations (BAO), see e.g. \cite{GilMarin2017}, or the luminosity distances to supernovae of Type Ia (SNIa) \cite{Scolnic2018}, just to mention some of them. However, in the last years and thanks mainly to the improvement of the observational facilities at our disposal, some cosmological tensions have arisen within the standard $\Lambda$CDM paradigm. Two of the most prominent ones are the so-called $H_0$ \cite{Verde2019,RiessReview2020,IntertwinedH0} and $\sigma_8$ tensions \cite{Macaulay2013,Nesseris2017,GomezValentSola2018,Intertwineds8}. In the following two paragraphs we comment on them, since they played an important role in \cite{ApJL2019,CQG2020} and hence also in the original motivation of this paper.

The Hubble-Lema\^itre parameter can be measured in an almost cosmology-independent way using the cosmic distance ladder method, see e.g. \cite{Robinson1985}. The SH0ES team has employed it to measure $H_0$ very accurately making use of the Hubble Space Telescope (HST). The last value reported by this collaboration, which has been obtained using also Gaia EDR3 parallaxes to calibrate an expanded list of Milky Way Cepheids in the first step of the ladder, reads $H_0=(73.2\pm 1.3)$ km/s/Mpc \cite{Riess2020}. This value is in $4.1\sigma$ tension with the one obtained by the Planck collaboration in the 2018 analysis of the CMB temperature, polarization and lensing data, under the assumption of the flat $\Lambda$CDM model, $H_0=(67.36\pm 0.54)$ km/s/Mpc \cite{Planck2018}. This tension could still be due to the presence of some systematic error in one or both data sets (see e.g. \cite{Efstathiou2020,Mortsell2021}), but at the moment there is no conclusive evidence of its existence, neither in Planck's \cite{Aylor2017,Addison2018} nor the HST \cite{Cardona2017,Zhang2017,Dhawan2018,Riess2019,Javanmardi2021} measurements. In addition, the tension is raised up to the $\sim 4.8\sigma$ level if the $H_0$ determination from the H0LICOW collaboration, $H_0=(73.3^{+1.7}_{-1.8})$ km/s/Mpc \cite{H0LICOW2020}, is combined with the SH0ES one. The former is obtained from the joint analysis of six gravitationally lensed quasars with measured time delays, and is independent of the latter. According to \cite{Denzel2020}, a lower tension (of $4.3\sigma$) is obtained when the systematic uncertainties of lensing degeneracies are duly taken into account in the strong lensing analysis. On the other hand, it is also very difficult to explain the discrepancy between the CMB and the distance ladder estimations of $H_0$ by a local underdensity. Due to the typical current values of the rms of mass fluctuations at scales of $8h^{-1}$ Mpc, $\sigma_8\sim 0.80-0.82$, it is very improbable that we live in one of such regions. The tension is only alleviated at the $\sim 3\sigma$ c.l. when the $\Lambda$CDM or its simplest extensions are considered \cite{Marra2013,Camarena2018}. See also \cite{Odderskov2016}. A much larger value of $\sigma_8$ would be needed to explain the $H_0$ tension in terms of the cosmic variance, but this, in turn, would exacerbate the $\sigma_8$ tension, so this cannot be considered a satisfactory solution.

The $\sigma_8$ tension is between the amount of large-scale structure (LSS) in the universe preferred by Planck's CMB data under the assumption of the $\Lambda$CDM model, and the one measured by several weak lensing (WL) and galaxy surveys. It is sometimes formulated in terms of the composite quantity $S_8=\sigma_8(\Omega_m^0/0.3)^{1/2}$, with $\Omega_m^0$ the normalized matter density parameter\footnote{See \cite{Ariel2020} for arguments against the use of $\sigma_8$ and related quantities as $S_8$.}. WL surveys are able to constrain $S_8$ at $\sim 5-6\%$ level \cite{Joudaki2018,Hildebrandt2020,Heymans2020}. For instance, according to KiDS+VIKING-450 cosmic shear data $S_8=0.716^{+0.043}_{-0.038}$ \cite{Hildebrandt2020}, which is in $2.7\sigma$ tension with the $\Lambda$CDM value obtained from the CMB TT,TE,EE+lowE+lensing analysis by Planck 2018, $S_8=0.832\pm 0.013$ ($\sigma_8=0.811\pm 0.006$) \cite{Planck2018}. The results by KiDS are completely compatible with those from DES \cite{Abbott2018}. By combining the KiDS-1000 WL results with BOSS and 2dFLenS data the authors of \cite{Heymans2020} have recently shown that it is possible to reduce by a factor $\sim 2$ the uncertainty on $S_8$, and also to write the tension only in terms of the $\sigma_8$ parameter, since BOSS puts tight constraints on $\Omega_m^0$. They report $S_8=0.766^{+0.020}_{-0.014}$ and $\sigma_8=0.760^{+0.021}_{-0.023}$, which are in $3.1\sigma$ and $2.2\sigma$ tension with Planck, respectively. Some initial hints of this tension were already found some years ago in data obtained from the analysis of redshift-space distortions (RSD) \cite{Macaulay2013}, and they are still there \cite{MNRAS2018,Nesseris2017,EPL2021}. These data are given in terms of the observable $f(z)\sigma_8(z)$, with $f=d\ln\,\delta_m/d\ln\,a$ being the growth rate and $\delta_m=\delta\rho_m/\rho_m$ the density contrast of matter fluctuations. RSD data tend to prefer also lower values of $\sigma_8$, especially when the bispectrum information is also exploited in the statistical analysis \cite{GilMarin2017,MNRAS2018,EPL2021,BorgesWands2020}.

If these tensions are not caused by the presence of any unaccounted systematic error in the data, they might be pointing to the need of some sort of new physics in the gravitational and/or matter sectors. Whatever this new physics is, it should alleviate the two tensions at a time or, at least, loosen one of them without worsening the other. This is a {\it golden rule} that is actually quite difficult to respect, see the discussion in \cite{CQG2020}. For instance, some models aiming to solve the $H_0$ tension tend to worsen the $\sigma_8$ one. This is the case of the early dark energy model studied in \cite{Poulin2019,Hill2020}, in which there is an enhancement of the Hubble rate in the pre-recombination epoch due to the presence of a scalar field in the universe with an associated constant and very large potential energy density that at $z\sim 3000$ decays as fast as (or even faster than) radiation. This excess of energy, which can be of about $\sim 7\%$, leads to a decrease of the sound horizon at the baryon-drag epoch, $r_s$, that has to be compensated by an increase of $H_0$ in order to keep the location of the first peak of the CMB temperature power spectrum untouched. The problem is that this also requires an increase of the current matter energy density to properly fit the overall data set and this, in turn, produces a larger growth of matter perturbations in the late-time universe. Hence, the $H_0$ tension in this model is alleviated, but at the expense of worsening the $\sigma_8$ one \cite{Hill2020}. Something similar is found in the early modified gravity model studied in \cite{Braglia2021}.

In \cite{ApJL2019,CQG2020} it was shown that it is possible to loosen both tensions at a time in the context of Brans-Dicke (BD) \cite{BransDicke1961,Brans1962,Dicke1962} cosmology with constant vacuum energy density (BD-$\Lambda$CDM)\footnote{Similar results are obtained for the Running Vacuum Model of Type II in \cite{EPL2021}. See therein for details.}, cf. also \cite{Sola2018,CruzSola2018}. For instance, these authors found $H_0=(71.30^{+0.80}_{-0.84})$ km/s/Mpc and $\sigma_8=0.789\pm 0.013$ when they confronted this model to a long string of cosmological observations, containing data from Planck 2018, the supernovae of Type (SnIa) from DES and the Pantheon compilation, cosmic chronometers, BAO, LSS, the prior on $H_0$ from SH0ES and the strong-lensing data from the H0LICOW collaboration. These values are fully compatible at $1.2\sigma$ and $1.1\sigma$ c.l. with those reported in \cite{Riess2020} and \cite{Heymans2020}, respectively, rendering the tensions harmless. The BD-$\Lambda$CDM model requires, though, values of the cosmological gravitational coupling $G\sim 4-9\%$ larger than the one measured on Earth, $G_N$, and also a mild evolution of $G$ throughout the expansion history\footnote{By forcing $G(z=0)=G_N$ it is also possible to alleviate the $H_0$ tension, but only to the $\sim 3.5\sigma$ level, and the $\sigma_8$ tension worsens in this scenario. See e.g. \cite{Ballardini2020}.}. Hence, it is obvious that in order to reconcile the theory with Solar System measurements and Cavendish-like experiments one needs to screen the modified gravity effects to recover standard GR locally. In this paper we study some extensions of the BD-$\Lambda$CDM model that could allow for such a screening. However, we find that only very tiny differences between the local and cosmological values of the gravitational coupling can be explained with these mechanisms, and argue that in order to explain these differences one needs to depart significantly from GR at some intermediate scale. These big departures break the validity of our weak-field limit approach, though, and should be studied in the future with care. If they are non-existent, our analysis jeopardizes the BD-$\Lambda$CDM model and some of its simplest extensions inside the Horndeski class \cite{Horndeski} as viable solutions to the cosmological tensions, and also points out the need of these models to produce $G(z=0)\simeq G_N$. This is the major result of this work. 

This paper is organized as follows. In Sec. \ref{sec:eBDLCDM} we present the action and field equations of our extended BD-$\Lambda$CDM model. In Sec. \ref{sec:screening} we discuss the Vainshtein and K-mouflage screening mechanisms in the context of this theory, by finding and analyzing the weak-field limit solutions for the scalar field under the assumption of a spherical static gravitational source with no internal pressure. We explain in Sec. \ref{sec:cubic} the limitations of the standard Vainshtein screening mechanism that is introduced by the derivative cubic interaction term $\propto \Box\psi\nabla_\mu\psi\nabla^\mu\psi$ in the action. These limitations are found even when the relative difference between the cosmological $G$ and $G_N$ is as small as $\mathcal{O}(10^{-9})$. In Sec. \ref{sec:kmouflage} we consider the derivative quartic self-interaction term $\propto (\nabla_\mu\psi\nabla^\mu\psi)^2$ in the action to study the K-mouflage screening, and show that it leads to the very same problem as in the Vainshtein mechanism. In Sec. \ref{sec:solutions} we allow for a non-constant $\wBD$ and show that in this case the screening is possible, but only if we allow for huge departures from GR at intermediate scales. Finally, we provide our conclusions in Sec. \ref{sec:conclusions}. Appendices \ref{sec:appendixA} and \ref{sec:AppendixB} complement some explanations given in the main text.


\newpage
\section{Extended Brans-Dicke theory. Action and field equations}\label{sec:eBDLCDM}

In this paper we study an extension of the BD-$\Lambda$CDM model \cite{CruzSola2018,ApJL2019,CQG2020,EPL2021}. BD-$\Lambda$CDM is built from the original BD-action \cite{BransDicke1961,Brans1962,Dicke1962}, but considering also the contribution of a cosmological constant term, $\rho_\Lambda$, which is in charge of triggering the late-time acceleration of the universe in this model, as it is in the standard GR-$\Lambda$CDM framework,

\begin{equation}\label{eq:LBDaction}
S_{\rm \Lambda BD}=\int d^4x \sqrt{-g}\left[\frac{1}{16\pi}\left(\psi R-\frac{\wBD}{\psi}\nabla^\mu\psi\nabla_\mu\psi\right)-\rho_\Lambda+\mathcal{L}_m(g_{\mu\nu},\chi_i)\right]\,.
\end{equation}
Here $\psi$ is a dynamical scalar field with dimensions of energy squared in natural units that is basically the inverse of the effective gravitational coupling at the action level, i.e. $\psi=1/G$. $\mathcal{L}_m$ is the Lagrangian density of the matter fields $\chi_i$, which accounts for the particles of the standard model and extensions of the latter needed to explain dark matter, massive neutrinos, etc. As in the original BD model, $\psi$ is non-minimally coupled to the Ricci scalar $R$, and its kinetic term is controlled by the BD-parameter $\wBD=const$.

In this work we extend the BD-$\Lambda$CDM action, as follows:

\begin{equation}\label{eq:action}
S_{\rm eBD}=S_{\rm \Lambda BD}+\int d^4x \frac{\sqrt{-g}}{16\pi} \left(f\,\Box\psi+ \theta\,\nabla^\nu\psi\nabla_\nu\psi\right)\nabla^\mu\psi\nabla_\mu\psi\,,
\end{equation}
where eBD refers to the ``extended Brans-Dicke'' theory that we are considering. The new piece of the action contains a cubic and a quartic self-interaction terms with constant couplings $f$ and $\theta$ with dimensions of energy to the power $-6$ and $-8$ in natural units, respectively. Apart from that, we also allow for a dependence of the BD-parameter on the scalar field, i.e. $\wBD=\wBD(\psi)$. These modifications of the original action \eqref{eq:LBDaction} do not alter the speed of propagation of gravitational waves, which is still $c^2_{gw}=1$, as in standard GR and BD theories, see e.g. \cite{Creminelli2017,Zumalacarregui2017}. By setting $f$ and $\theta$ to 0 and $\wBD=const.$ we recover the BD-$\Lambda$CDM model \eqref{eq:LBDaction} studied in \cite{ApJL2019,CQG2020,CruzSola2018,EPL2021}, of course.

We use for convenience the dimensionless scalar field
\begin{equation}
\varphi\equiv G_N\psi\,,
\end{equation} 
with $G_N=1/m^2_{\rm Pl}$ the local value of the gravitational coupling and $m_{\rm Pl}\simeq 1.22\times 10^{19}$ GeV the Planck mass\footnote{We use natural units, i.e. $\hbar=c=1$, but still keep the explicit dependence on $G_N$ in the formulas.}. At the action level, we can define the effective coupling

\begin{equation}\label{eq:effGaction}
G(\varphi) = \frac{G_N}{\varphi}\,,
\end{equation}
which does not coincide in general with the effective gravitational strength between two test masses, $G_{\rm eff}$, see e.g. \cite{CQG2020} or Appendix \ref{sec:AppendixB}. The variation of the action  \eqref{eq:action} with respect to the metric field $g_{\mu\nu}$ and the scalar field $\varphi$ leads to the modified Einstein and Klein-Gordon equations, which read, respectively\footnote{For the geometrical quantities, we use the metric $(-, +,+,+ )$, Riemann tensor $R^\lambda_{\,\,\,\,\mu \nu \sigma} = \partial_\nu \, \Gamma^\lambda_{\,\,\mu\sigma} + \Gamma^\rho_{\,\, \mu\sigma} \, \Gamma^\lambda_{\,\, \rho\nu} - (\nu \leftrightarrow \sigma)$, Ricci tensor $R_{\mu\nu} = R^\lambda_{\,\,\,\,\mu \lambda \nu}$, and Ricci scalar $R = g^{\mu\nu} R_{\mu\nu}$. These correspond to the $(+, +, +)$ conventions in the famous classification by Misner, Thorn and Wheeler \cite{Misner1973}.}, 

\begin{align}\label{eq:Eeq}
&\varphi G_{\mu\nu}+g_{\mu\nu}\left[\Box\varphi+\frac{\wBD}{2\varphi}(\nabla\varphi)^2\right]-\nabla_\mu\nabla_\nu\varphi-\frac{\wBD}{\varphi}\nabla_\mu\varphi\nabla_\nu\varphi+\frac{g_{\mu\nu}}{2}\frac{f}{G_N^2}\nabla^\alpha\varphi\nabla_\alpha(\nabla\varphi)^2\\
&+\frac{f}{G_N^2}\Box\varphi \nabla_\mu\varphi\nabla_\nu\varphi-\frac{f}{G_N^2}\nabla_\mu(\nabla\varphi)^2\nabla_\nu\varphi+\frac{2\theta}{G_N^3}(\nabla\varphi)^2\left[\nabla_\mu\varphi\nabla_\nu\varphi-\frac{g_{\mu\nu}}{4} (\nabla\varphi)^2\right]= 8\pi G_N (T_{\mu\nu}-\rho_\Lambda g_{\mu\nu})\,,\nonumber
\end{align}

\begin{align}\label{eq:KG1}
0=R&+\frac{1}{\varphi}\left(\wBD^\prime-\frac{\wBD}{\varphi}\right)(\nabla\varphi)^2+\frac{2\wBD}{\varphi}\Box\varphi\\
&+\frac{f}{G_N^2}\Box(\nabla\varphi)^2-\frac{2f}{G_N^2}\nabla_\mu\left(\Box\varphi\nabla^\mu\varphi\right)-\frac{4\theta}{G_N^3}\nabla_\nu\left[\nabla^\nu\varphi(\nabla\varphi)^2\right]\,,\nonumber
\end{align}     
where $(\nabla\varphi)^2\equiv\nabla_\mu\varphi\nabla^\mu\varphi$, $\Box\equiv \nabla_\mu\nabla^\mu$, and the prime in \eqref{eq:KG1} denotes a derivative with respect to $\varphi$. $G_{\mu\nu}=R_{\mu\nu}-(1/2)g_{\mu\nu}R$ is the Einstein tensor and $T_{\mu\nu}=-(2/\sqrt{-g})\delta S_m/\delta g^{\mu\nu}$ the total energy-momentum tensor of the relativistic and non-relativistic matter fields, with  $T=g^{\mu\nu}T_{\mu\nu}$. 

We can take the trace of equation \eqref{eq:Eeq} to remove $R$ in \eqref{eq:KG1} and write the equation for the scalar field in an alternative (more standard) form,

\begin{align}\label{eq:KG2}
(3+2\wBD)&\Box\varphi=8\pi G_N (T-4\rho_\Lambda)-\wBD^\prime(\nabla\varphi)^2+\frac{4\theta\varphi}{G_N^3}\nabla_\nu\left[\nabla^\nu\varphi(\nabla\varphi)^2\right]\\
&+\frac{2f\varphi}{G_N^2}\left[(\Box\varphi)^2-(\nabla_\mu\nabla_\nu\varphi)\left(\nabla^\mu\nabla^\nu\varphi+\frac{1}{\varphi}\nabla^\mu\varphi\nabla^\nu\varphi\right)-\frac{1}{2\varphi}\Box\varphi(\nabla\varphi)^2-R_{\mu\nu}\nabla^\mu\varphi\nabla^\nu\varphi\right]\,.\nonumber
\end{align}
It is evident from this equation that in the limit $\wBD\to\infty$ there exists a trivial solution $\varphi=const$. The dynamics of the scalar field is turned off in this limit and, in particular, we can retrieve standard GR if we set $\varphi=1$, i.e. $G=const.=G_N$. In general, though, some evolution of $\varphi$ is allowed by the data. This evolution is due to a departure of $1/\wBD$ from 0. When the dynamics is very mild, as it happens e.g. in the cosmological scenario for typical values of the $\wBD$ parameter \cite{LiWuChen2013,AvilezSkordis2014,CruzSola2018,ApJL2019,CQG2020,EPL2021,Umilta2015,Ballardini2016,Ballardini2020}, the non-linear terms coming from the cubic and quartic self-interactions that appear in the {\it rhs} of equation \eqref{eq:KG2} are expected to be negligible, although this will ultimately depend, of course, on the particular values of the couplings $f$ and $\theta$, since the latter determine the scale at which non-linearities start playing a role. If, on the contrary, the dynamics of the scalar field is important, non-linearities can eventually be dominant in equation \eqref{eq:KG2} and $\varphi=const.$ is then again a solution, leading to the screening of the Brans-Dicke effects \cite{SilvaKoyama2009,SchmidtHuLima2010,Koyama2020} through the Vainshtein \cite{Vainshtein} or K-mouflage \cite{Babichev2009} mechanisms. The speed at which the scalar field attains the non-linear regime is controlled by $\wBD$. When $\wBD$ approaches $-3/2$ the transition times/lengths between the linear and non-linear regimes shorten and the transition happens faster, eventually becoming instantaneous in the limit $\wBD\to -3/2$. These preliminary comments will be duly qualified and extended in the next section, where we discuss in detail the screening of the eBD model \eqref{eq:action}.


\newpage
\section{Screening in the eBD theory}\label{sec:screening}

As already mentioned in the introduction, some works in the literature that study the BD-$\Lambda$CDM model \cite{ApJL2019,CQG2020,EPL2021} find hints in favor of a departure from GR at cosmological scales, with an effective gravitational coupling $G(\varphi)=G_N/\varphi$ mildly varying with the expansion and central values $\sim 4-9\%$ larger than the one measured on Earth, $G_N$\,\footnote{The statistical confidence level of these signals depends on the concrete cosmological data set employed in the fitting analysis, cf. e.g. Tables 3 and 4 of \cite{CQG2020}.}. The dynamics of the scalar field at the cosmological level is constrained to be pretty small, due to the large values required for $\wBD^{(c)}$ or, equivalently, small values required for its inverse, $1/\wBD^{(c)}\equiv\epsilon^{(c)}_{\rm BD}\lesssim\mathcal{O}(10^{-3})$. Here, and from now on, we use the superscript $(c)$ to indicate the cosmological nature of these quantities. In \cite{AvilezSkordis2014} the authors found results fully compatible with GR using CMB data from WMAP+SPT and a prior on the Hubble parameter from the HST. Their results still allow, though, for departures of $\sim 10\%$ in the cosmological value of $G$ from $G_N$ at $1\sigma$ c.l. (see also \cite{Salvado2015}) and similar constraints for $\epsilon^{(c)}_{\rm BD}$ to those obtained in \cite{CruzSola2018,ApJL2019,CQG2020,EPL2021} with more recent data. Nevertheless, despite the smallness of $\epsilon^{(c)}_{\rm BD}$ there is still plenty of room for values of this parameter much larger than those allowed by measurements in the Solar System. The Cassini mission \cite{Bertotti2003} constrains the local (Solar System) value of $\eBD$ to be $\eBD^{(l)}=(-2.1\pm 2.3)\times 10^{-5}$, where the superscript $(l)$ is employed to denote local quantities. This upper bound is two orders of magnitude lower than the one obtained for $\eBD^{(c)}$. On the other hand, $\varphi^{(l)}=1$ locally, since the gravitational coupling is given by Newton's constant $G_N$, whereas one finds values of $\varphi^{(c)}(z=0)$ in the range  $0.9-1$ to be fully compatible with cosmological observations. Hence, in order to reconcile the central values of $G$ and $\eBD$ inferred from cosmological analyses with those measured locally, it is mandatory to implement a screening mechanism. The Vainshtein screening \cite{Vainshtein} of the modified gravity effects of Brans-Dicke theory has been already studied in the literature \cite{SilvaKoyama2009,SchmidtHuLima2010,Koyama2020} making use of the cubic interaction term that appears in \eqref{eq:action}. We will show in Sec. \ref{sec:cubic}, though, that this new piece in the action is not sufficient to explain differences between $\varphi^{(c)}$ and $\varphi^{(l)}$ of order $\mathcal{O}(0.1)$, as the ones we are interested in. A similar result is found in Sec. \ref{sec:kmouflage} making use of the quartic interaction term of \eqref{eq:action}, which is responsible of the K-mouflage screening \cite{Babichev2009}.

In Sec. \ref{sec:cubic} we study the effect of the cubic interaction term by still considering a constant $\wBD$, and discuss its limitations concerning the screening. In Sec. \ref{sec:kmouflage} we obtain analogous results using the quartic term. In  Sec. \ref{sec:solutions} we allow for the evolution of $\wBD=\wBD(\varphi)$ and study some possible forms of this function. All of them seem to lead to too big departures from GR at intermediate scales. 


\subsection{Vainshtein screening}\label{sec:cubic}

Let us consider the eBD model \eqref{eq:action}, with $\wBD=const.$ and $\theta=0$. We want to obtain the scalar field profile created by a spherical static mass embedded in empty space, and in the weak-field limit. This is equivalent to embed the mass in an expanding Friedmann-Lema\^itre-Robertson-Walker (FLRW) universe when we are interested in time scales for the system much lower than the inverse of the Hubble function, i.e. $H^{-1}$. Therefore, we neglect the effect of the low background matter energy density and the cosmological constant. We expand the metric around Minkowski, $\eta_{\mu\nu}$, and the scalar field around its background cosmological value, $\varphi_0\equiv\varphi^{(c)}$,

\begin{equation}\label{eq:pert}
g_{\mu\nu}=\eta_{\mu\nu}+h_{\mu\nu} \qquad ;\qquad \varphi=\varphi_0+\delta\varphi\,,
\end{equation}
with $|\eta_{\mu\nu}|\gg |h_{\mu\nu}|$ and $|\varphi_0|\gg |\delta\varphi|$. The truncated perturbative expansions will work at any point outside sources with a radius $\mathcal{R}$ much larger than their associated Schwarzschild radius, $r_{Sch}(M)=2G_N M$, with $M$ being the total mass of the gravitating body. In order to study the effect of the cubic interaction term appearing in the action \eqref{eq:action} we need to keep in equation \eqref{eq:KG2} at least the contributions of second order in perturbation theory,

\begin{align}\label{eq:eqA}                                                                                                                         
(\eta^{\mu\nu}+h^{\mu\nu})\partial_\mu\partial_\nu\delta\varphi+\eta^{\mu\nu}&\Gamma^\kappa_{\mu\nu}(h)\partial_\kappa\delta\varphi=\frac{-8\pi G_N\rho}{3+2\wBD}\\
&+\frac{2f\varphi_0}{G_N^2(3+2\wBD)}\left[(\eta^{\mu\nu}\partial_\mu\partial_\nu\delta\varphi)^2-(\partial_\mu\partial_\nu\delta\varphi)(\partial^\mu\partial^\nu\delta\varphi)\right]\,,\nonumber
\end{align}
where here we have used the energy-momentum tensor of a pressureless perfect fluid at rest (dust), so $T=-\rho$. Notice that the Christoffel symbols $\Gamma^\kappa_{\mu\nu}(h)$ are already of first order in perturbations, see Appendix \ref{sec:appendixA}. In the {\it lhs} of equation \eqref{eq:eqA} we are allowed to neglect the terms of first order in $h_{\mu\nu}$, so we are left with no dependence on the perturbed metric in this equation. Thus, in the static limit the latter reads,

\begin{equation}\label{eq:ijequa}
\partial^i\partial_i\delta\varphi=\frac{-8\pi G_N\rho}{3+2\wBD}+\frac{2f\varphi_0}{G_N^2(3+2\wBD)}\left[(\partial^i\partial_i\delta\varphi)^2-(\partial_i\partial_j\delta\varphi)(\partial^i\partial^j\delta\varphi)\right]\,,
\end{equation}
with $\partial^i\equiv \eta^{ij}\partial_j=\delta^{ij}\partial_j$ and, therefore, $\partial^i\partial_i=\nabla^2$ the Laplace operator. As we are considering a static spherical mass distribution, i.e. $\rho=\rho(r)$, the latter can be rewritten in spherical coordinates as follows,

\begin{equation}
\frac{1}{r^2}\frac{d}{dr}\left(r^2\frac{d\delta\varphi}{dr}\right)=\frac{-8\pi G_N\rho(r)}{3+2\wBD}+\frac{4f\varphi_0}{G_N^2(3+2\wBD)}\left[\frac{1}{r^2}\left(\frac{d\delta\varphi}{dr}\right)^2+\frac{2}{r}\frac{d\delta\varphi}{dr}\frac{d^2\delta\varphi}{dr^2}\right]\,.
\end{equation}
The integration of this differential equation is straightforward, and yields,

\begin{equation}\label{eq:soeq}
r^2\frac{d\delta\varphi}{dr}=\frac{-2G_NM(r)}{3+2\wBD}+\frac{4f\varphi_0}{G_N^2(3+2\wBD)}r\left(\frac{d\delta\varphi}{dr}\right)^2\,, 
\end{equation}
which allows us to isolate the derivative of $\delta\varphi$ with respect to $r$ in a simple way. For values of $r>\mathcal{R}$ we can substitute $M(r)$ by $M$ and obtain,

\begin{equation}\label{eq:der1}
\frac{d\delta\varphi}{dr} = \frac{rG_N^2(3+2\wBD)}{8f\varphi_0}\left[1-\sqrt{\frac{32Mf\varphi_0}{r^3(3+2\wBD)^2G_N}}\right]\,.
\end{equation}
The other solution to the second order equation \eqref{eq:soeq} is not considered, since it does not lead to the desired Brans-Dicke solution at large distances. It is useful and physically enlightening to define the following two characteristic lengths,
           
\begin{equation}\label{eq:lengthscales}
r_c=\left[\frac{8f\varphi_0}{G_N^2(3+2\wBD)}\right]^{1/2}\qquad ; \qquad r_V=\left[\frac{32Mf\varphi_0}{G_N(3+2\wBD)^2}\right]^{1/3}\,,
\end{equation} 
and rewrite \eqref{eq:der1} in a more compact way,

\begin{equation}\label{eq:der2}
\frac{d\delta\varphi}{dr} =\frac{r}{r_c^2}\left[1-\sqrt{1+\frac{r_V^3}{r^3}}\right]\,.
\end{equation} 
Both lengths, $r_c$ and $r_V$, grow with the coupling $f$. The first controls the power of the screening mechanism, whereas the second, known as Vainshtein radius, sets the border between the screened and unscreened regions. Let us study now the solutions of \eqref{eq:der2} in the limit of distances much lower and much larger than the Vainshtein radius.

For $r\gg |r_V|$ one recovers the pure Brans-Dicke solution \eqref{eq:deltavarphi}, 

\begin{equation}\label{eq:BDsolution}
\delta\varphi(r)=\frac{2G_NM}{(3+2\wBD) r}\,,
\end{equation}
regardless of the sign and value of $f$, which means that the far field solution is not sensitive to the details of the screening mechanism, as expected, since non-linearities play no role there. The sign of the constant $(3+2\wBD)$ determines the sign of the derivative of $\varphi$ farther away from $r_V$. The scalar field decreases with $r$ if $(3+2\wBD)>0$, and increases if $(3+2\wBD)<0$. This is important. If $\varphi_0<1$ we need $(3+2\wBD)>0$ in order to match the cosmological value of the scalar field with the local one, since $\varphi(r\to 0)\equiv \varphi_{Scr}=1$; and if $\varphi_0>1$ we need $(3+2\wBD)<0$. Other combinations do not allow us to explain the local observations in the context of the screening mechanism under study, with $\wBD^\prime=0$. 

For $r\ll |r_V|$,

\begin{equation}\label{eq:lowr}
\frac{d\delta\varphi}{dr}=-\left(\frac{3+2\wBD}{|3+2\wBD|}\right)\left(\frac{G_N^3M}{2f\varphi_0 r}\right)^{1/2}\longrightarrow \delta\varphi(r)=-\left(\frac{3+2\wBD}{|3+2\wBD|}\right)\left(\frac{2G_N^3Mr}{f\varphi_0}\right)^{1/2}+\delta\varphi_{Scr}.\,,
\end{equation}
with $\delta\varphi_{Scr}$ an integration constant that depends on all the parameters of the theory. We will determine it later on. This solution tells us that the BD effects get suppressed at sufficiently small values of $r$, below the Vainshtein radius, since $\varphi$ tends to a constant when $r\to 0$. Therefore, we can in principle retrieve GR by forcing $\varphi_0+\delta\varphi_{Scr}=1$. In addition, in order the scalar field to take real values inside the screened region we have to demand $f\geq 0$, of course. This automatically ensures the positivity of $r_V$, which makes sense because it is a real physical length; $r_c$, instead, can still be imaginary if $(3+2\wBD)<0$. This is not a problem, since only its square enters \eqref{eq:der2}. The sign of $(3+2\wBD)$ fixes basically the sign of $d\delta\varphi/dr$ outside and inside the Vainshtein radius, as we have already discussed before. Hence, $r_c$ should not be thought of as a physical length because it can be imaginary. However the inverse of its absolute value is actually a physical energy scale (in natural units) that determines the threshold from which non-linearities become relevant. The bigger is $f$ the lower is the latter and the sooner non-linearities take the control of equation \eqref{eq:ijequa}.  

The general solution to equation \eqref{eq:der2} for $r\geq\mathcal{R}$ is well known \cite{SilvaKoyama2009,SchmidtHuLima2010,Koyama2020},

\begin{equation}\label{eq:intermediate}
\delta\varphi(r)=\left(\frac{r_V}{r_c}\right)^2\int_{\infty}^{r/r_V}dx\,x\,[1-\sqrt{1+x^{-3}}]=\left(\frac{r_V}{r_c}\right)^2 g\left(\frac{r}{r_V}\right)\,,
\end{equation}
with

\begin{equation}\label{eq:gfunction}
g(x) = \frac{x^2}{2}\left[1-\,_2F_1\left(-\frac{1}{2},-\frac{2}{3},\frac{1}{3},-x^{-3}\right)\right]\,,
\end{equation}
and $_2F_1$ the ordinary hypergeometric function. For illustrative purposes we plot $g\left(r/r_V\right)$ in Fig. 1. It is easy to check that in the limits $r\gg r_V$ and $r\ll r_V$ we recover from \eqref{eq:intermediate} the solutions \eqref{eq:BDsolution} and \eqref{eq:lowr}, respectively, as previously discussed. The formula \eqref{eq:intermediate} allows us to link the local and cosmological values of the scalar field. Assuming that $\varphi^{(l)}=1$ is measured at distances much lower than the Vainshtein radius, where we recover standard GR, and making use of the fact that $g(x)\to 2.103$ when $x\to 0$ we find,

\begin{equation}
1=\varphi_0+2.103\left(\frac{r_V}{r_c}\right)^2\,.
\end{equation}
Multiplying this expression by $r_V$ and using \eqref{eq:lengthscales} we get a simple relation between the cosmological value of the scalar field, the Vainshtein and Schwarzschild radii, and $\wBD$,

\begin{figure}[t!]
\begin{center}
\label{fig:gfunction}
\includegraphics[width=6in, height=2in]{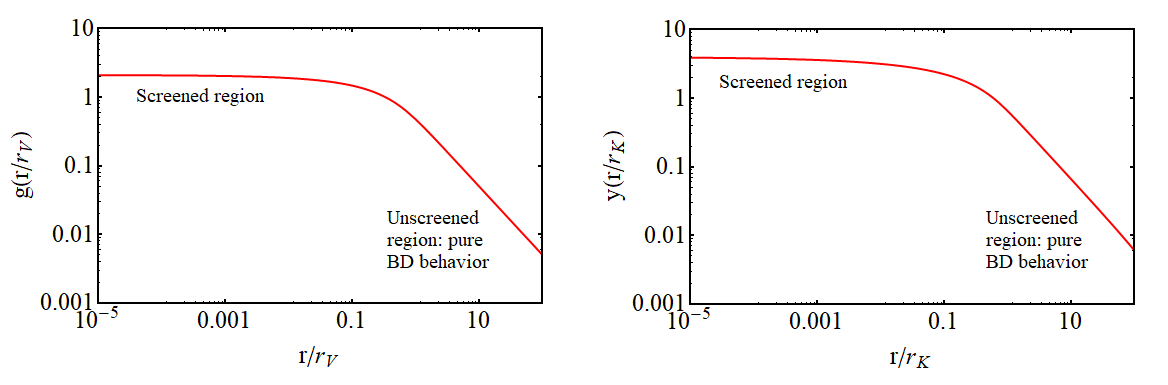}
\caption{\scriptsize {\it Left plot:} Shape of the function $g(x)$ \eqref{eq:gfunction} as a function of $x=r/r_V$; {\it Right plot:} The same for $y(x)$ \eqref{eq:yfunction} as a function of $x=r/r_K$.}
\end{center}
\end{figure}

\begin{equation}\label{eq:problemScr}
(1-\varphi_0)\,r_V= 4.206\,\frac{r_{Sch}(M)}{3+2\wBD}\,.
\end{equation}
In order to be consistent, we have to demand $r_V\gg \mathcal{R}\gg r_{Sch}(M)$, which implies 

\begin{equation}
\frac{4.206}{(1-\varphi_0)(3+2\wBD)}\gg 1\,.
\end{equation}
It is evident that for values of $1-\varphi_0\sim\mathcal{O}(0.1)$ and $(3+2\wBD)^{-1}\approx\epsilon^{(c)}_{BD}/2\sim 10^{-3}$ as the ones allowed (and in some cases even preferred) by cosmological observations this condition is not fulfilled. The reason is that the Brans-Dicke running of the scalar-field is too slow due to the smallness of $\eBD^{(c)}$. With such values of $\eBD^{(c)}$, we can choose a coupling $f$ to accomodate large differences between $\varphi^{(l)}=1$ and $\varphi^{(c)}$, but the screening is achieved at very small values of the radius $r$, since non-linearities cannot be produced at larger distances. The Vainshtein radius is excessively small and, actually, smaller than the Schwarzschild radius. This even falls out of the validity range of our solution \eqref{eq:intermediate} for various reasons, e.g. because at $r\sim r_{Sch}$ the weak-field limit approximation does not hold any more. Alternatively, if we choose $f$ to generate reasonable values of the Vainshtein radius, i.e. values of $r_V$ larger than the radius $\mathcal{R}$ of the source, we are only able to explain very tiny differences $1-\varphi_0\lesssim\mathcal{O}(10^{-8})$ for the Milky Way or $1-\varphi_0\lesssim\mathcal{O}(10^{-12})$ for the Solar System. This can be clearly seen in Fig. 2, and similar conclusions are obtained if we consider bigger structures, e.g. a typical galaxy cluster of mass $M\sim 10^{14}\,M_{\odot}$ and radius $\mathcal{R}\sim 1$ Mpc. Thus, it is obvious from these arguments that the screening mechanism under consideration is not able to match the local measurements of $G$ with the $\sim 10\%$ departures of $\varphi_0$ allowed by the cosmological data \cite{AvilezSkordis2014,Salvado2015,ApJL2019,CQG2020,EPL2021}. Higher order Galileon terms in the action, e.g. $(\nabla^\mu\varphi\nabla_\mu\varphi)^2\Box\varphi$ or $(\nabla^\mu\varphi\nabla_\mu\varphi)^3\Box\varphi$, also allow for the screening, as expected, but suffer from the very same problems found in the cubic case. Their activation happens once non-linearities start to be significant. This is only possible for tiny values of the Vainshtein radius when one considers the typical values of $\epsilon_{\rm BD}$ allowed by the cosmological data and $1-\varphi_0\sim\mathcal{O}(0.1)$. We show in the next section that the same problem is encountered in the context of the K-mouflage mechanism.

\begin{figure}[t!]
\begin{center}
\label{fig:limitcubic}
\includegraphics[width=6.5in, height=2.5in]{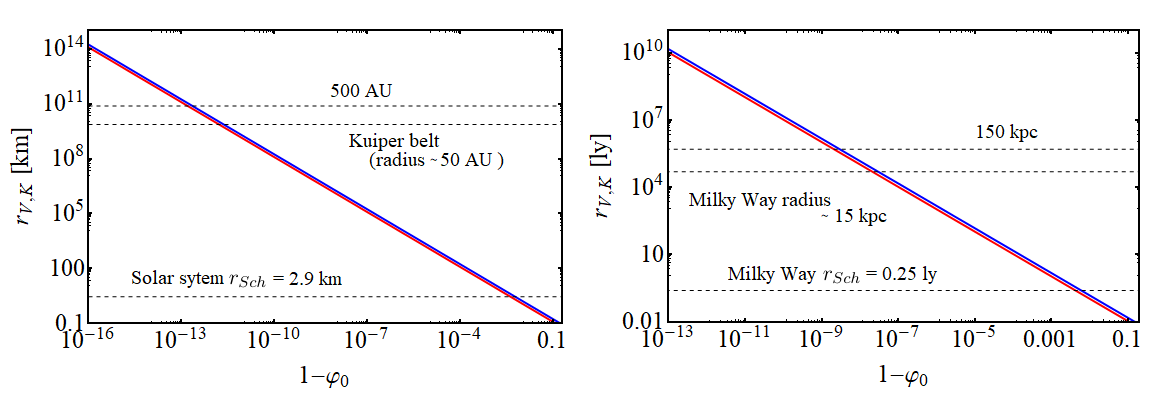}
\caption{\scriptsize {\it Left plot:} In red, the size of the Vainshtein radius associated to the Solar Sytem as a function of $1-\varphi_{0}$, obtained upon setting $\eBD^{(c)}=2\cdot 10^{-3}$ and using formula \eqref{eq:problemScr}, which assumes a perfect spherical source, of course. In blue, the K-mouflage radius, using \eqref{eq:probKmou}. We also indicate in black dashed lines the corresponding Schwarzschild radius, the radius of the Kuiper belt ($\sim 50$ AU), and ten times the radius of the Kuiper belt; {\it Right plot:} The same, but for the Milky Way. In black dashed lines we show again the corresponding Schwarzschild radius, together with its physical radius (of about 15 kpc), and a distance equal to ten times the latter. In both cases, ones needs values of $1-\varphi_0$ extremely close to 0 to produce reasonable estimates of $r_V$ and $r_K$. Larger values of $\eBD^{(c)}$ than the one employed to generate these plots are in conflict with cosmological observations, and lower values worsen even more the problem under discussion. See the main text for further explanations.}
\end{center}
\end{figure}

\newpage
\subsection{K-mouflage screening}\label{sec:kmouflage}

Now we consider the action \eqref{eq:action} with $\wBD=const.$ and $f=0$. The resulting model is of K-mouflage type \cite{Babichev2009}. The equation for the scalar field in the weak-field limit reads,

\begin{equation}\label{eq:kmou1}
\partial^i\partial_i\delta\varphi = \frac{-8\pi G_N}{3+2\wBD}+\frac{4\theta\varphi_0}{G_N^3(3+2\wBD)}[(\partial_j\delta\varphi)(\partial^j\delta\varphi)\partial_i\partial^i\delta\varphi+2(\partial^i\delta\varphi)(\partial^j\delta\varphi)\partial_i\partial_j\delta\varphi]\,.
\end{equation}
Making use of 

\begin{equation}
2(\partial^i\delta\varphi)(\partial^j\delta\varphi)\partial_i\partial_j\delta\varphi=\partial_{i}[\partial^i\delta\varphi(\partial_j\delta\varphi)(\partial^j\delta\varphi)]-(\partial_j\delta\varphi)(\partial^j\delta\varphi)\partial_i\partial^i\delta\varphi
\end{equation}
in \eqref{eq:kmou1} we obtain,

\begin{equation}
\partial^i\partial_i\delta\varphi = \frac{-8\pi G_N}{3+2\wBD}+\frac{4\theta\varphi_0}{G_N^3(3+2\wBD)}\partial_{i}[\partial^i\delta\varphi(\partial_j\delta\varphi)(\partial^j\delta\varphi)]\,.
\end{equation}
This equation can be easily integrated, assuming also spherical symmetry, and using the Gauss theorem in the last term of the {\it rhs}. For $r>\mathcal{R}$,

\begin{equation}\label{eq:kmou2}
\frac{d\delta\varphi}{dr}= \frac{-2G_NM}{(3+2\wBD)r^2}+\frac{4\theta\varphi_0}{G_N^3(3+2\wBD)}\left(\frac{d\delta\varphi}{dr}\right)^3\,.
\end{equation}
This is a cubic equation for the derivative $d\delta\varphi/dr$. The form of its solution depends on the sign of the discriminant. One can easily see that the only physically meaningful solution is found when the discriminant is positive with $\theta(3+2\wBD)<0$\footnote{The discriminant cannot be zero $\forall{r}$ if $M\ne 0$ and $3+2\wBD\ne 0$. On the other hand, the discriminant can only be negative at sufficiently large distances if $\theta(3+2\wBD)>0$, but the corresponding solution in that limit does not behave like in pure BD, so it has to be discarded.}. In this case there are only one real and two imaginary solutions. We are interested in the real one, of course, which is given by Cardano's formula. We can define the following quantities with units of length, in an analogous way to what we did in Sec. \ref{sec:cubic} for the Vainshtein screening, 

\begin{equation}\label{eq:relKmou}
\tilde{r}_c=\frac{4\theta\varphi_0}{G_N^4 M} \qquad ;\qquad r_K=\left(\frac{-108\,\theta\,\varphi_0 M^2}{G_N(3+2\wBD)^3}\right)^{1/4}\,.
\end{equation}
Notice that $r_K$ is always real and positive due to the condition mentioned before, $\theta(3+2\wBD)<0$. Equation \eqref{eq:kmou2} can be rewritten as follows,

\begin{equation}\label{eq:kmou3}
\frac{d\delta\varphi}{dr}=\frac{1}{\tilde{r}^{1/3}_cr^{2/3}}\left[\left(1+\sqrt{1+\left(\frac{r}{r_K}\right)^4}\right)^{1/3}-\left(-1+\sqrt{1+\left(\frac{r}{r_K}\right)^4}\right)^{1/3}\right]\,,
\end{equation}
and one can see that $\tilde{r}_c$ and $r_K$ in this model play the same role as $r_c$ and $r_V$ in the model of the previous section. When $r\gg r_K$ we recover the pure Brans-Dicke solution \eqref{eq:BDsolution}, whereas for $r\ll r_K$ we find,

\begin{equation}
\frac{d\delta\varphi}{dr}=\left(\frac{G_N^4 M}{2\theta\varphi_0 r^2}\right)^{1/3}\longrightarrow \delta\varphi(r)=3\left(\frac{G_N^4 Mr}{2\theta\varphi_0 }\right)^{1/3}+\delta\varphi_{Scr}\,.
\end{equation}
As in the Vainshtein mechanism, $\delta\varphi\to \delta\varphi_{Scr}^+$ when $(3+2\wBD)<0$ (or, equivalently, when $\theta>0$) and $\delta\varphi\to \delta\varphi_{Scr}^-$ when $(3+2\wBD)>0$. The solution to equation \eqref{eq:kmou3} valid for $\forall{r}>\mathcal{R}$ is given by

\begin{equation}
\delta\varphi(r)=-\left(\frac{r_K}{\tilde{r}_c}\right)^{1/3} y\left(\frac{r}{r_K}\right),
\end{equation}
with

\begin{equation}\label{eq:yfunction}                                            
y\left(z\right) = \int_{\infty}^{z}dx\,x^{-2/3}\,\left[\left(-1+\sqrt{1+x^4}\right)^{1/3}-\left(1+\sqrt{1+x^4}\right)^{1/3}\right]\,,
\end{equation}
see the right plot in Fig. 1. It is evident that the Vainhstein and K-mouflage screening mechanisms lead to very similar profiles for the scalar field and, therefore, we expect to find the very same problem that we described in Sec. \ref{sec:cubic}. One can show that in this case, 

\begin{equation}\label{eq:probKmou}
(1-\varphi_0)\,r_K = 5.976\,\frac{r_{Sch}(M)}{3+2\wBD}\,,
\end{equation}
where here we have made use of the fact that $y(x)\to 3.984$ when $x\to 0$, and also of the relations \eqref{eq:relKmou}. It is not possible to obtain reasonable values of $r_K$ when $1-\varphi_0\sim \mathcal{O}(0.1)$ and $\epsilon^{(c)}_{BD}\sim 10^{-3}$, cf. Fig. 2. By combining the Vainshtein and K-mouflage mechanisms we can have a {\it double screening} framework, in which the one that dominates depends on the mass of the source, allowing for different phenomenology at different scales \cite{Gratia2016}. However, it is obvious that by combining the Vainshtein and K-mouflage mechanisms we cannot solve our problem.

We study in Sec. \ref{sec:solutions} what happens if we consider a variable $\wBD=\wBD(\varphi)$.


\subsection{Screening with a varying $\wBD$}\label{sec:solutions}

As discussed in the previous sections, the low values of $\eBD^{(c)}$ allowed by the cosmological data \cite{AvilezSkordis2014,ApJL2019,CQG2020,EPL2021,CruzSola2018} (which are still two orders of magnitude larger than those measured in the Solar System \cite{Bertotti2003}) lead to a too mild spatial variation of the scalar field. There are no values of the couplings $f$ nor $\theta$ able to produce at a time differences $1-\varphi_0\sim \mathcal{O}(0.1)$ and reasonable (high enough) values for the Vainshtein/K-mouflage radii. We have seen that both mechanisms work in a very similar way. Hence, keeping them in the action at a time does not improve the situation. To solve this problem we are forced to modify the action studied in Secs. \ref{sec:cubic} and \ref{sec:kmouflage}. Equations \eqref{eq:problemScr} and \eqref{eq:probKmou} tell us that by taking a value of $\wBD$ close enough to $-3/2$ we could make the job, in principle. This would generate big spatial gradients of the scalar field that would allow it to reach the non-linear regime faster (at larger distances), increasing in this way the Vainshtein/K-mouflage radii. The problem is that values $\wBD\sim -3/2$ are clearly not allowed neither by the cosmological data nor measurements carried out in the Solar System. Still, there could be regions in the Universe, i.e. at some intermediate scales, in which gravity has not been directly tested yet, where these large gradients could exist in principle. It is important to remark that the weak-field limit approximation would break in these regions. To explicitly show that values $\wBD\sim -3/2$ are needed we have to consider a non-constant $\wBD$. By choosing its form appropriately this function can also allow us to flatten the shape of $\varphi(r)$ at small distances, achieving in practice the desired screening without the need of using cubic or quartic self-interaction terms. Thus, for the sake of simplicity we deactivate these interactions by setting $f=0$ and $\theta=0$. In the weak-field limit, equation \eqref{eq:KG2} can be written as follows, 

\begin{equation}
\partial^i\partial_i\delta\varphi=\frac{8\pi G_N}{3+2\wBD}T-\frac{\wBD^\prime}{3+2\wBD}\partial_i\delta\varphi\partial^i\delta\varphi\,.
\end{equation}
If we consider, again, spherical symmetry,

\begin{equation}
\frac{1}{r^2}\frac{d}{dr}\left(r^2\frac{d\delta\varphi}{dr}\right)=\frac{-8\pi G_N\rho(r)}{3+2\wBD}-\frac{\wBD^\prime}{3+2\wBD}\left(\frac{d\delta\varphi}{dr}\right)^2\,.
\end{equation}
For $r>\mathcal{R}$ we have $\rho(r)=0$, so 

\begin{equation}
\frac{1}{r^2}\frac{d}{dr}\left(r^2\frac{d\delta\varphi}{dr}\right)=-\frac{\wBD^\prime}{3+2\wBD}\left(\frac{d\delta\varphi}{dr}\right)^2\,,
\end{equation}
which can be easily solved to obtain the gradient of the scalar field as a function of $\wBD$, 

\begin{equation}\label{eq:eqwBD}
r^2\frac{d\delta\varphi}{dr} = \frac{C}{\sqrt{3+2\wBD}}\,,
\end{equation}
with $C$ an integration constant that can be set for instance by demanding $d\delta\varphi/dr$ to match the pure Brans-Dicke solution \eqref{eq:deltavarphi} sufficiently far away from the source, at some radius $r_*$, 

\begin{figure}[t!]
\begin{center}
\label{fig:limitcubic}
\includegraphics[width=6.6in, height=2.in]{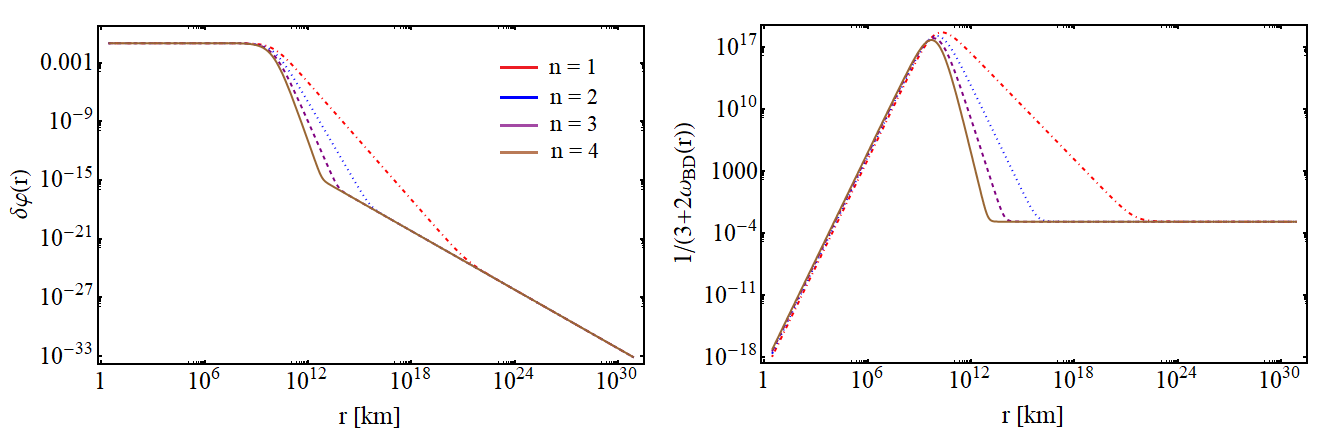}
\caption{\scriptsize Plots of the scalar field $\delta\varphi(r)$ \eqref{eq:PhenoField} and its associated BD function $\wBD(r)$ \eqref{eq:eqwBD} for the Solar System ($r_{Sch}=2.9$ km). We set $\varphi_0=0.9$ ($\delta\varphi_{Scr}=0.1$), $r_V\sim 50$ AU, and study different values of the parameter $n$.}
\end{center}
\end{figure}

\begin{equation}
C= r_*^2\frac{d\varphi}{dr}\Big\rvert_{r_*}\sqrt{3+2\wBD^{(c)}}= \frac{-r_{Sch}}{\sqrt{3+2\wBD^{(c)}}}\,.
\end{equation}
We can specify $\wBD(\varphi)$ and compute $\delta\varphi(r)$ integrating equation \eqref{eq:eqwBD}. This allows us to obtain $\wBD(r)$; we can also start by choosing a particular $\wBD(r)$ to obtain $\delta\varphi(r)$ and then invert it to obtain $r(\delta\varphi)$ and finally $\wBD(\varphi)$; or what is even more useful for our purposes, we can select the shape $\delta\varphi(r)$ that we desire and compute $\wBD(r)$, and then find $\wBD(\varphi)$. This is a simple way to check what shapes of $\wBD(\varphi)$ are needed to generate the correct phenomenology, and e.g. estimate the lowest value of $3+2\wBD$ needed, together with the distance from the source at which the largest deviation from GR happens. Thus, expression \eqref{eq:eqwBD} can help us to determine whether Brans-Dicke model with a varying $\wBD$ can perform the efficient screening we are looking for or not.

For instance, we can explore the following phenomenological expression for $\delta\varphi(r)$, 

\begin{equation}\label{eq:PhenoField}
\delta\varphi(r) = \frac{1}{(r+r_V)^{n+1}}\left[\delta\varphi_{\rm Scr}r_V^{n+1}+\frac{r_{Sch}r^n}{3+2\wBD}\right]\,,
\end{equation}
with $r_V$ being the characteristic radius at which the screening happens and $n$ a positive integer number that controls the slope of the scalar field between the pure Brans-Dicke ($r\gg r_V$) and fully screened ($r\ll r_V$) regions. In Fig. 3 we show the shape of $\wBD(r)$ obtained for the Solar System ($r_{Sch}=2.9$ km), considering $\varphi_0=0.9$ ($\delta\varphi_{Scr}=0.1$), $r_V\sim 50$ AU and different values of $n$. As expected, in order to match $10\%$ deviations of the gravitational coupling $G$ from $G_N$ it is necessary to fine-tune a lot the value of $\wBD$ to $-3/2$ in some region of space. Although in this limit our weak-field approximation clearly breaks down, we expect this to lead to large deviations from GR. In addition, we would need to understand why there is this fine-tuning in the value of $\wBD$, and explain why this happens when $\delta\varphi$ is as small as $\sim 10^{-15}$ or so (cf. again Fig. 3). This is very unnatural. Other forms of $\delta\varphi(r)$ can also be studied, but the problem remains in essence, since it is due to the need of large gradients of the scalar field. The introduction of a non-constant potential for the scalar field does not solve the problem neither, since regardless of its shape it has to generate large gradients of $\delta\varphi(r)$, which in turn are the origin of the aforementioned problems. Hence, we do not expect the latter to be solved with the aid of chameleon \cite{chameleon} or symmetron \cite{symmetron} mechanisms. They are of course able to screen Brans-Dicke effects in dense enough environments, but they are not capable of explaining the large relative differences between the cosmological and local values of the gravitational coupling that are needed to loosen the cosmological tensions.


\section{Conclusions}\label{sec:conclusions}

Some recent cosmological studies have found some significant evidence in favor of the Brans-Dicke model with a constant vacuum energy density (BD-$\Lambda$CDM) when it is constrained under a very complete set of cosmological data \cite{ApJL2019,CQG2020,EPL2021}. This model of modified gravity is able to alleviate the $H_0$ and $\sigma_8$ tensions at a time if the cosmological value of the effective gravitational coupling entering the action evolves mildly with the expansion of the universe and is about $4-9\%$ larger than the value measured on Earth. These values fall also inside the validity range of other previous analyses in the literature \cite{AvilezSkordis2014,Salvado2015}. Some sort of screening mechanism is needed to match the cosmological and local values of $G$. In this paper we have studied several extensions of the BD-$\Lambda$CDM model that are contained inside the general family of Horndeski theories \cite{Horndeski}. More concretely, we have computed the weak-field limit profile of the Brans-Dicke scalar field generated by a static massive source with no internal pressure, assuming two different types of screening mechanisms: Vainshtein and K-mouflage. We have seen that they are only able to explain very small differences between the value of $G$ that enters the Friedmann equation and the one that is measured locally at the Solar System scale. Allowing for a field-dependent $\wBD$ can potentially help to alleviate the problem, but only if there are huge departures from GR at intermediate scales, which should be studied numerically with care since they break the weak-field approximation considered in this work. These departures are most probably non-existent. If so, {\it our analysis suggests that the current value of the cosmological gravitational coupling can only be extremely close to $G_N$ in the context of the scalar-tensor theories of gravity that reduce to BD-$\Lambda$CDM at large scales, and this can be used as a boundary condition in any cosmological study dealing with these theories. There is still plenty of room, though, for a non-negligible time evolution of $G$ (both, at local and cosmological scales) throughout the cosmic history, governed by an $|\epsilon_{\rm BD}^{(c)}|=1/|\wBD^{(c)}|\lesssim 10^{-3}$}. The local value could in principle follow very closely the cosmological one, although this should be studied in more detail, probably in the context of McVittie metric \cite{McVittie}. Local constraints on the time evolution of $G$ would apply in this case \cite{Babichev2011}.

It is important to remark that in this work we have not focused on alternative scalar-tensor theories of gravity with Vainshtein or K-mouflage mechanisms that depart from the BD-$\Lambda$CDM at cosmological scales, as e.g. \cite{Benevento2019}, so our conclusions do not affect these theories. Finally, it is also worth to mention that in \cite{EPL2021} the authors showed that some running vacuum models (RVMs) with an induced variation of the gravitational coupling (Ricci RVM of Type II) exhibit a very similar phenomenology to the BD-$\Lambda$CDM model when confronted to a rich set of cosmological data and, in particular, they are also able to alleviate the cosmological tensions. The derivatives of $G$ might not enter Einstein field equations in these RVMs. This could allow to have an efficient screening mechanism without generating the large departures from GR found in the framework of the BD-$\Lambda$CDM model. 

\vspace{0.5cm}

{\bf Acknowledgements}
The authors are grateful to Prof. Luca Amendola and Prof. Joan Sol\`a Peracaula for their feedback. AGV is funded by the Deutsche Forschungsgemeinschaft (DFG) - Project number 415335479. 


\appendix

\section{Some geometrical quantities up to linear order in perturbations}\label{sec:appendixA}

In this short appendix we provide for completeness the formulas of some geometrical quantities up to first order in the perturbed metric $h_{\mu\nu}$, cf. \eqref{eq:pert}. The Christoffel symbols, Ricci tensor, Ricci scalar and Einstein tensor read, respectively:
                               
\begin{equation}\label{eq:Christoffel1}
\Gamma^\alpha_{\beta\kappa}(h) = \frac{\eta^{\alpha\mu}}{2}\left(h_{\mu\beta,\kappa}+h_{\mu\kappa,\beta}-h_{\beta\kappa,\mu}\right)\,,
\end{equation}

\begin{equation}
R_{\mu\nu}(h)=\frac{1}{2}\left[\partial^\alpha\partial_\nu h_{\alpha\mu}+\partial^\alpha\partial_\mu h_{\alpha\nu}-\partial_{\mu}\partial_\nu h-\eta^{\alpha\beta}\partial_\alpha\partial_\beta h_{\mu\nu}\right]\,,
\end{equation}

\begin{equation}
R(h)=\partial^\alpha\partial^\mu h_{\alpha\mu}-\eta^{\alpha\mu}\partial_\alpha\partial_\mu h\,,
\end{equation}

\begin{equation}\label{eq:ET1}
G_{\mu\nu}(h)=\frac{1}{2}\left[\partial^\alpha\partial_\nu h_{\alpha\mu}+\partial^\alpha\partial_\mu h_{\alpha\nu}-\partial_{\mu}\partial_\nu h-\eta^{\alpha\beta}\partial_\alpha\partial_\beta h_{\mu\nu}+\eta_{\mu\nu}(\eta^{\alpha\beta}\partial_\alpha\partial_\beta h-\partial^\alpha\partial^\beta h_{\alpha\beta})\right]\,.
\end{equation}


\section{Weak-field limit in the original Brans-Dicke theory}\label{sec:AppendixB}

In the pure Brans-Dicke theory we have the following equations ruling the behavior of the perturbed scalar and metric fields in the weak-field limit:

\begin{equation}\label{eq:BDvarphi}
\partial^i\partial_i\delta\varphi=\frac{-8\pi G_N\rho}{3+2\wBD}\,,
\end{equation}
\begin{equation}\label{eq:eqB}
\varphi_0 G_{\mu\nu}(h)+\eta_{\mu\nu}\partial^\alpha\partial_\alpha\delta\varphi-\partial_\mu\partial_\nu\delta\varphi = 8\pi G_N T_{\mu\nu}\,,
\end{equation}
where $G_{\mu\nu}(h)$ is given by formula \eqref{eq:ET1}. It is sufficient to consider here perturbations up to linear order. These equations are coupled, but one can remove the dependence on $\delta\varphi$ in \eqref{eq:eqB} by performing the following change \cite{Koyama2020}: 

\begin{equation}\label{eq:change}
h_{\mu\nu}=H_{\mu\nu}-\eta_{\mu\nu}\frac{\delta\varphi}{\varphi_0}\,.
\end{equation}
Upon substitution of \eqref{eq:change} in \eqref{eq:eqB} we obtain,

\begin{equation}\label{eq:eqB2}
G_{\mu\nu}(H) = \frac{8\pi G_N}{\varphi_0} T_{\mu\nu}\,.
\end{equation}
From the 0i component of \eqref{eq:eqB2} we get $H_{0i}(r)=0$, and by summing its 00 component and its trace,

\begin{equation}\label{eq:H00}
-\nabla^2 H_{00} = \frac{8\pi G_N}{\varphi_0}\rho \longrightarrow H_{00}(r)=\frac{2G_NM}{\varphi_0 r}\,.
\end{equation}
Equation \eqref{eq:BDvarphi} can be also solved to find

\begin{equation}\label{eq:deltavarphi}
\delta\varphi(r)=\frac{2G_NM}{(3+2\wBD) r}\,.
\end{equation}
The gravitational acceleration of a free-falling non-relativistic particle is given by $a^i=-\Gamma^{i}_{00}(h)=\partial^i h_{00}/2$, where $h_{00}$ can be directly computed using \eqref{eq:H00} and \eqref{eq:deltavarphi} in \eqref{eq:change}. This allows us to infer the effective gravitational coupling in the pure Brans-Dicke theory, which is constant and reads \cite{BransDicke1961},

\begin{equation}
G_{\rm eff}= G(\varphi_0)\left(\frac{4+2\wBD}{3+2\wBD}\right)\,,
\end{equation} 
with $G(\varphi)$ given by \eqref{eq:effGaction}.


\end{document}